\documentclass[preprint,12pt]{elsarticle}
\usepackage{comment}
\usepackage{xcolor}
\usepackage{array}
\usepackage{arydshln}
\usepackage{amsmath}
\usepackage{subcaption}
\usepackage{gensymb}
\usepackage{cancel}
\usepackage{makecell}
\usepackage{graphicx,booktabs}
\usepackage{multicol}
\usepackage{makecell}

\usepackage[margin=2.5cm]{geometry}

\usepackage[separate-uncertainty=true]{siunitx}
\DeclareSIUnit\torr{torr}

\journal{Nuclear Materials and Energy}

\begin{document}

\title{Particle control via cryopumping and its impact on the edge plasma profiles of Alcator C-Mod}

\author[1]{M.A. Miller\corref{cor1}}
\ead{millerma@mit.edu}
\author[1]{J.W. Hughes}
\author[2]{S. Mordijck}
\author[1]{M. Wigram}
\author[1]{J. Dunsmore}
\author[3]{R. Reksoatmodjo}
\author[4]{R.S. Wilcox}

\cortext[cor1]{Corresponding author}

\address[1]{MIT Plasma Science and Fusion Center, Cambridge, MA 02139, USA}
\address[2]{William \& Mary, Williamsburg, VA 23188, USA}
\address[3]{Lawrence Livermore National Laboratory, Livermore, CA 94550}
\address[4]{Oak Ridge National Laboratory, P.O. Box 2008, Oak Ridge, TN 37831}

\date{June 2024}

\begin{abstract}
    At the high $n_{e}$ proposed for high-field fusion reactors, it is uncertain whether ionization, as opposed to plasma transport, will be most influential in determining $n_{e}$ at the pedestal and separatrix. A database of Alcator C-Mod discharges is analyzed to evaluate the impact of source modification via cryopumping. The database contains similarly-shaped H-modes at fixed $I_{p} =$ 0.8 MA and $B_{t} =$ 5.4 T, spanning a large range in $P_\mathrm{net}$ and ionization. Measurements from an edge Thomson Scattering system are combined with those from a midplane-viewing Ly$_{\alpha}$ camera to evaluate changes to $n_{e}$ and $T_{e}$ in response to changes to ionization rates, $S_\mathrm{ion}$. $n_{e}^\mathrm{sep}$ and $T_{e}^\mathrm{ped}$ are found to be most sensitive to changes to $S_\mathrm{ion}^\mathrm{sep}$, as opposed to $n_{e}^\mathrm{ped}$ and $T_{e}^\mathrm{sep}$. Dimensionless quantities, namely $\alpha_\mathrm{MHD}$ and $\nu^{*}$, are found to regulate attainable pedestal values. Select discharges at different values of $P_\mathrm{net}$ and in different pumping configurations are analyzed further using SOLPS-ITER. It is determined that changes to plasma transport coefficients are required to self-consistently model both plasma and neutral edge dynamics. Pumping is found to modify the poloidal distribution of atomic neutral density, $n_{0}$, along the separatrix, increasing $n_{0}$ at the active X-point. Opaqueness to neutrals from high $n_{e}$ in the divertor is found to play a role in mediating neutral penetration lengths and hence, the poloidal distribution of neutrals along the separatrix. Pumped discharges thus require a larger particle diffusion coefficient than that inferred purely from 1D experimental profiles at the outer midplane.
\end{abstract}

\maketitle

\section{Introduction}
\label{sec:intro}

Successful operation of a fusion reactor will require the ability to regulate the confined plasma's particle content \cite{houlberg_density_1994}. Since fusion power scales as the square of the plasma density, $n$ \cite{lawson_criteria_1957}, the ability to control this parameter is highly desirable from a performance point of view. Once in the burning plasma regime, where heating may be largely dominated by alpha particles, particle control may be one of the few levers remaining to control fusion power generation. The details of the 1D (let alone 2D or 3D) distribution of $n$ are the result of a number of different physical processes, resulting from the interplay between plasma, neutrals, and material surfaces. It is, however, important to consider how much influence a controller might have on the overall particle content in the plasma.

From the point of view of performance and the ability to integrate a high confinement scenario with a power exhaust solution, it is important to answer this question at two particular locations in the plasma edge. The first of these is specifically relevant for the high-confinement mode (H-mode), where pressure gradients steepen significantly resulting from the strengthening of a shear layer in the edge and local reduction in the cross-field transport \cite{asdex_team_h-mode_1989, burrell_effects_1997}. The density, $n$, at the top of this steep-gradient region, the so-called ``pedestal", given by $n_\mathrm{ped}$, is of particular importance. Fusion performance in the H-mode depends to a large extent on the pedestal pressure, $p_\mathrm{ped}$, due to stiff transport in the core. This yields rather large sensitivity of core density and temperature, $T$, to the boundary conditions at the top of the pedestal \cite{kotschenreuther_quantitative_1995, greenwald_h_1997, kinsey_iter_2011, frassinetti_global_2017, rodriguez-fernandez_predictions_2020}. $n_\mathrm{ped}$ and $p_\mathrm{ped}$ are themselves related, and their exact relationship depends strongly on the type of H-mode present \cite{hughes_pedestal_2013, faitsch_analysis_2023}. For the typical ELMy H-mode, for example, their relationship depends on the details of pedestal stability, in particular whether the pedestal is on the peeling or the ballooning branch of the peeling-ballooning stability boundary \cite{snyder_characterization_2004}. 

The second location important to control is the density at the separatrix, $n_\mathrm{sep}$. This quantity is considered very important for power handling \cite{kallenbach_neutral_2019, henderson_parameter_2021}. Since dissipative processes scale strongly with $n^{2}$ \cite{Moulton_Lengyel_2021}, the ability for a divertor to handle plasma fluxes from the fusion-producing core is a strong function of $n_\mathrm{sep}$. Further, this parameter has been linked to a number of important boundaries for plasma operation. The first of these is the transition between the low and high confinement mode, the L-H transition \cite{eich_separatrix_2021}. Once in H-mode, evidence exists that $n_\mathrm{sep}$ (as well as $T_\mathrm{sep}$) may play a large role in determining the specific type of H-mode produced \cite{faitsch_analysis_2023}. Finally, recent work has linked the edge density, particularly that at the separatrix, to the well known density limit \cite{manz_power_2023}. Control over $n_\mathrm{sep}$ may thus also be important for avoiding density-limit induced plasma disruptions.

Section \ref{sec:experiment} begins by outlining the parameters of the database studied, drawing on previous analysis showing the impact of $P_\mathrm{net}$ as well as the pump state on particle inventory and confinement. It introduces the diagnostics used, one which measures the electron density and temperature, $n_{e}$ and $T_{e}$, and the other which measures neutral emission. Section \ref{sec:ionization} looks at how $S_\mathrm{ion}$ changes with pumping and how $S_\mathrm{ion}$ itself modifies edge $n_{e}$ and $T_{e}$. The next section then considers how these changes at both the separatrix and pedestal top occur as a function of dimensionless quantities. Section \ref{sec:solps} then seeks to study the importance of 2D effects on determining both heat and particle sources. Section \ref{sec:poloidal_transport} looks in detail about how poloidal distributions might affect particle fluxes and thus, particle transport in the pedestal. Finally, Section \ref{sec:conclusions} makes concluding remarks, discussing the findings and proposing next steps.

\section{Experiments to determine the impact of pumping}
\label{sec:experiment}

\begin{figure}[b!]
\centering
\includegraphics[width=0.3\columnwidth]{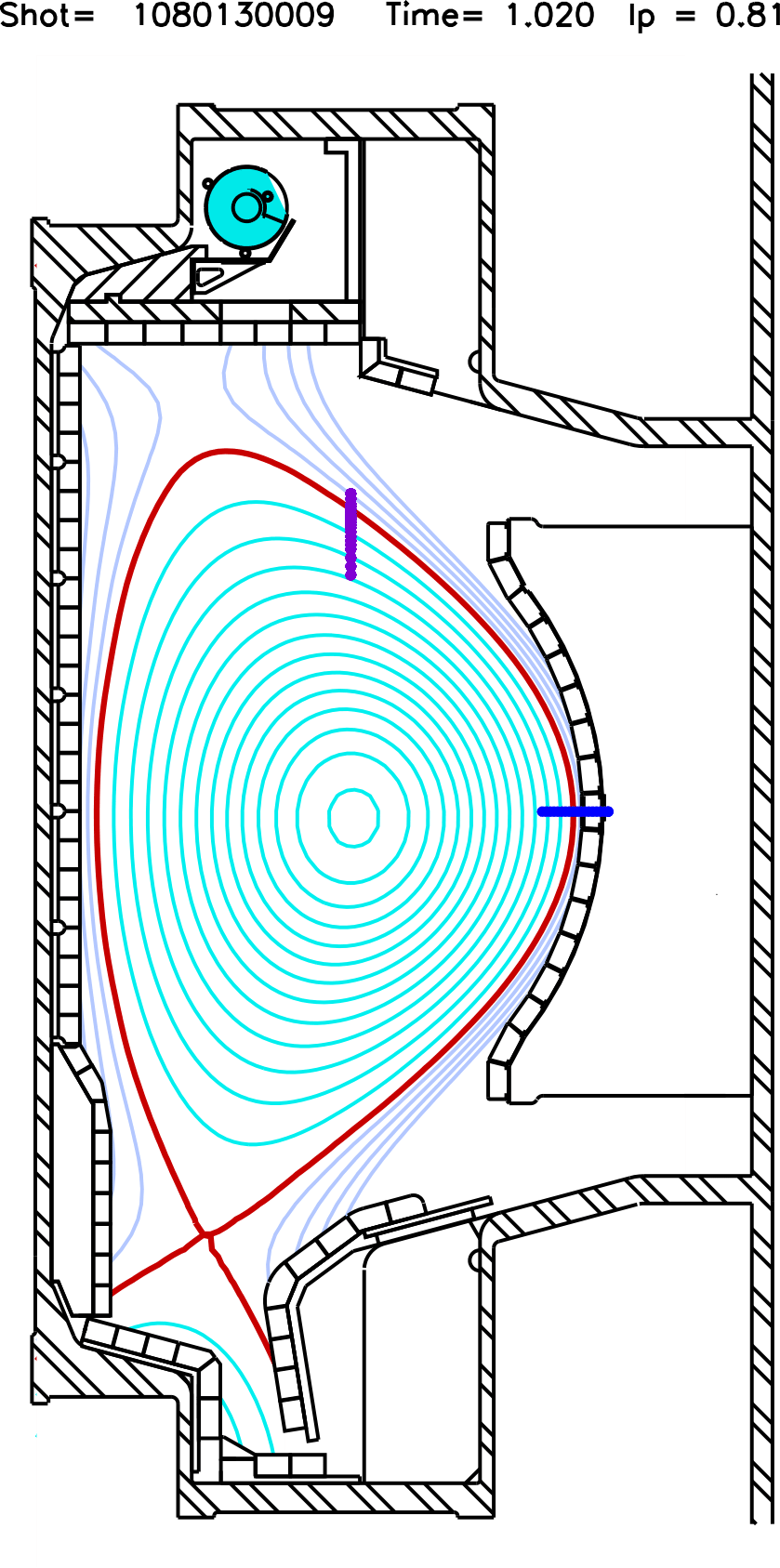}
\caption{Typical magnetic equilibrium for pumped discharges. The cryopump volume is shown in turquoise at the top of the diagram. Shown also are two diagnostics: the ETS (purple), viewing near the plasma crown, and LYMID (bright blue), viewing near the outer midplane.}
\label{fig:xs_pump}
\end{figure}

\begin{figure}
\centering
\includegraphics[width=\columnwidth]{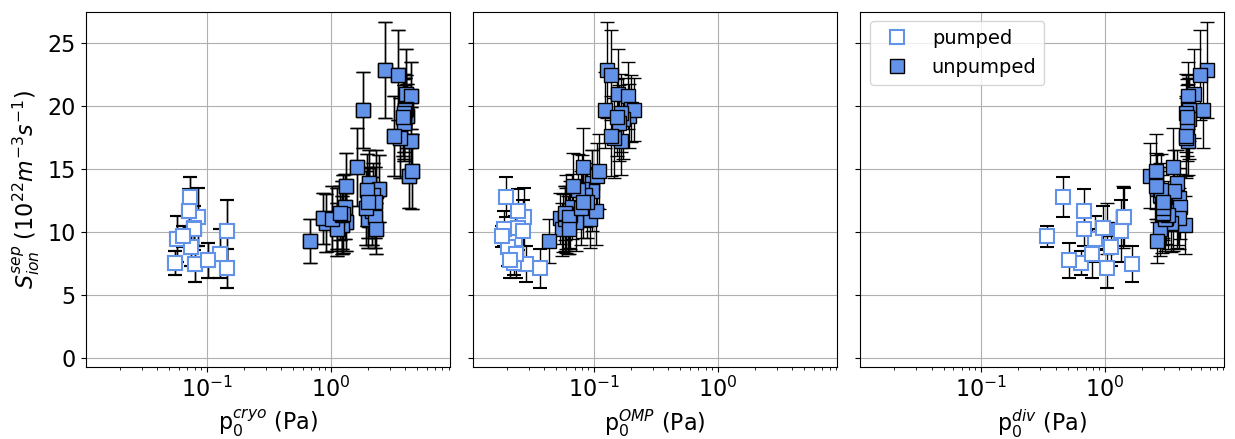}
\caption{Ionization rate at the separatrix plotted against neutral pressure measured above the cryopump volume (left), behind the limiter at the OMP (center), and below the outer strike point (right), for both the pumped (open squares) and unpumped (closed squares).}
\label{fig:p0_Sion}
\end{figure}

This work examines two sets of matched H-mode discharges from Alcator C-Mod. The first had no active particle control while in the H-mode. Fueling occurred exclusively in the L-mode phase, and the H-mode was fueled entirely via recycling fluxes. The second, however, made use of a cryopump designed to lower the particle inventory. This pump, installed in 2007 was housed in the upper divertor chamber, as shown in turquoise in Figure \ref{fig:xs_pump}. It achieved pumping speeds as high as 9,600 L/s \cite{brian_pump_2007}. Despite a measurable decrease in neutral pressures as a result of pumping (Figure \ref{fig:p0_Sion}), no additional gas puff was requested via the feedback fueling system during the H-mode phase, similarly to when the pump was turned off.

All of these discharges were performed at fixed $B_{t}$ = 5.4 T and $I_{p}$ = 0.8 MA. The plasmas were similarly-shaped (elongation, $\kappa \approx 1.7$, and lower triangularity, $\delta_{l} \approx 0.54$) and all lower single null, with the $\nabla B$-drift direction pointing towards the active lower null, i.e. the direction favorable for H-mode access. The distance between the primary and secondary separatrices, $\Delta R_\mathrm{sep}$, ranged between 3 - 5 mm. Figure \ref{fig:xs_pump} shows a typical equilibrium of one of these discharges. Note that the cryopump is located in the upper divertor, near the upper outer leg, but these discharges are all in lower null. The effectiveness of cryopumping is largely tied to the amount of particle flux arriving at the pump duct. Since $\Delta R_\mathrm{sep}$ was not large, even though the active null was far from the pump duct, the presence of the secondary null in the vacuum vessel allowed a non-negligible amount of particle flux to be removed by the cryopump.

One of the two primary diagnostics used in this work is the edge Thomson scattering (ETS) diagnostic. ETS measured both electron density ($n_{e}$) and electron temperature ($T_{e}$), spanning 3 cm of the plasma edge. It was located near the plasma crown and could diagnose $n_{e}$ and $T_{e}$ with order millimeter resolution when mapped to the midplane \cite{hughes_high-resolution_2001}. A second diagnostic, the LYMID camera, measured Ly$_{\alpha}$ emission at the midplane. It recorded line-integrated brightness measurements from each of its viewing chords as a function of their tangency radius. Combined with measurements from ETS and a collisional-radiative model, it allows reconstruction of radial profiles of neutral atomic density, $n_{0}$, ionization rate, $S_\mathrm{ion}$, and particle flux, $\Gamma_{D}$ are constructed from a Ly$_{\alpha}$ camera called LYMID. Details of the calculation of quantities derived from these two diagnostics are explained in detail in \cite{miller_collisionality_2024}.

\begin{figure}
\centering
\includegraphics[width=0.5\columnwidth]{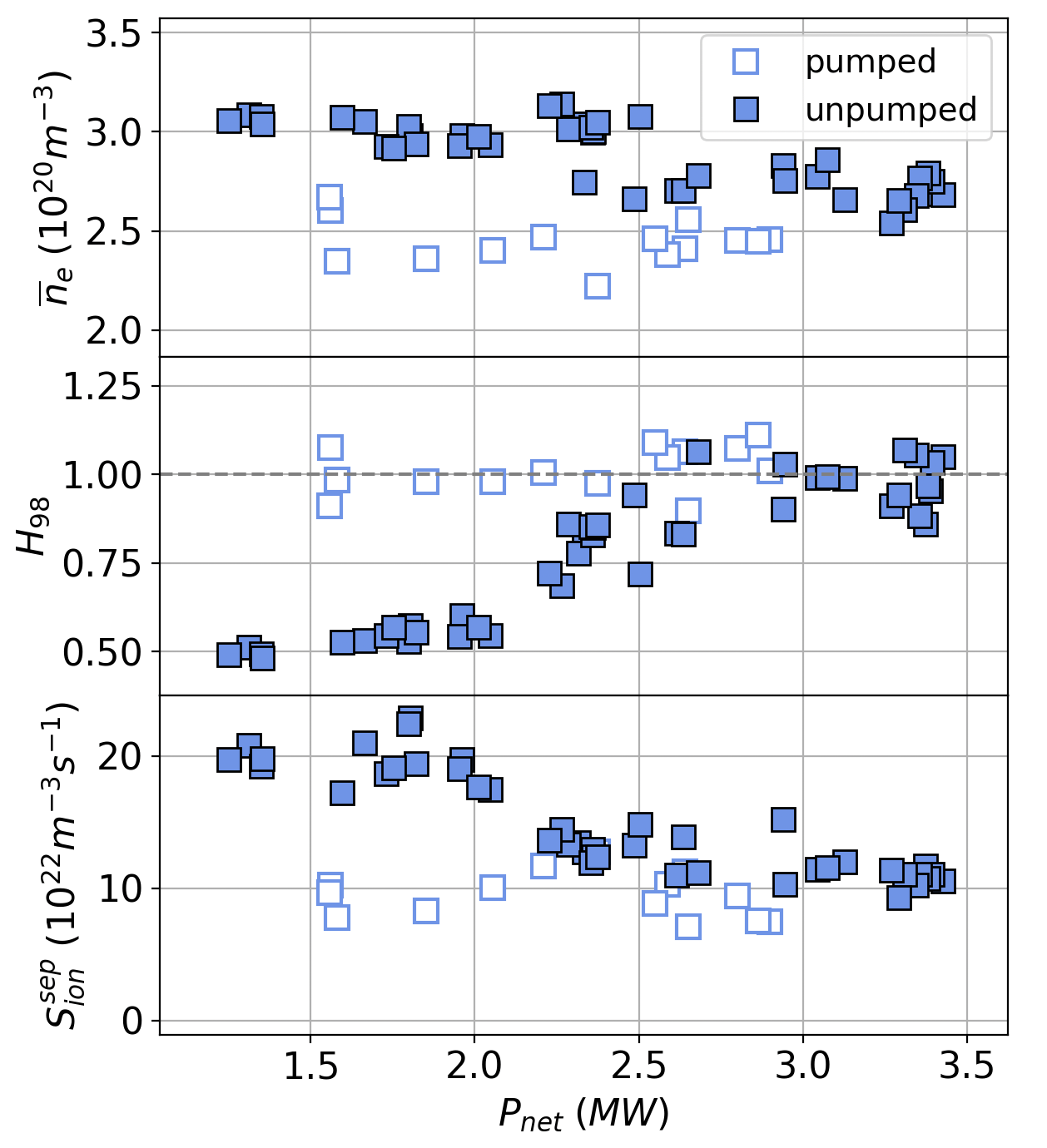}
\caption{Line-averaged density (top panel), normalized confinement, (middle panel), and separatrix ionization (bottom panel) against $P_\mathrm{net}$, for both pumped (open symbols) and unpumped (closed symbols) discharges. Pumping slightly reduces $\overline{n}_{e}$ and maintains $H_{98} = 1$, while keeping $S_\mathrm{ion}^\mathrm{sep}$ low, even at low $P_\mathrm{net}$.}
\label{fig:jerry_pnet}
\end{figure}

Both sets of experiments featured an effective scan in the power crossing the separatrix, $P_\mathrm{net}$, calculated as $P_\mathrm{net} = P_\mathrm{loss} - P_\mathrm{rad}$. Here, $P_\mathrm{loss}$ is the loss power, calculated using $P_\mathrm{loss} = P_\mathrm{tot} - dW/dt$, and $P_\mathrm{rad}$ is the power radiated in the core. $P_\mathrm{tot}$ is the total power, taken to be the sum of the ion cyclotron range of frequencies (ICRF) power ($P_\mathrm{RF}$) and the ohmic power ($P_\mathrm{oh}$), and $dW/dt$ is the time derivative of the stored energy, $W$. $P_\mathrm{net}$ in these discharges ranged between $1.3 - 3.4$ MW. In the unpumped set of experiments, shown with closed blue symbols in Figure \ref{fig:jerry_pnet}, a clear change to both the line-integrated density, $\overline{n}_{e}$, and the normalized confinement, $H_{98}$, was observed as $P_\mathrm{net}$ dropped below a critical value, $P_\mathrm{net}^\mathrm{crit} \approx 2.3$ MW. The same figure shows that when the cryopump was turned on (open symbols), $\overline{n}_{e}$ was reduced by $\sim$15\%. Though there is some scatter in the measurement at low $P_\mathrm{net}$, there is no large decrease in confinement below the same value of $P_\mathrm{net}^\mathrm{crit}$. Importantly, $H_{98}$ remained near unity below $P_\mathrm{net}^\mathrm{crit}$.

A separate analysis examining the unpumped discharges only found that $S_\mathrm{ion}$ throughout the edge was highly sensitive to changes in $P_\mathrm{net}$ \cite{miller_collisionality_2024}. These rates were inferred from emission of the Ly$_{\alpha}$ transition of neutral deuterium, collected by a toroidally-viewing camera at the outer midplane (OMP). As $P_\mathrm{net}$ dropped below $P_\mathrm{net}^\mathrm{crit}$, $S_\mathrm{ion}^\mathrm{sep}$ grew rapidly. With higher ionization $n_{e}^\mathrm{sep}$ also increased, but $n_{e}^\mathrm{ped}$ was close to constant. This insensitivity of $n_{e}^\mathrm{ped}$ to changes in $P_\mathrm{net}$, or equivalently $S_\mathrm{ion}^\mathrm{sep}$, was linked to a rise in mid-pedestal particle transport, $D_\mathrm{eff}$, consistent with a fixed density gradient, $\nabla n_{e}$, but a significantly degraded the temperature gradient, $\nabla T_{e}$. This thus reduced the pressure  gradient, $\nabla p_{e}$, which ultimately reduced $H_{98}$ \cite{miller_collisionality_2024}. Figure \ref{fig:jerry_pnet} implies that pumping might enable one to avoid whatever driving transport mechanism was responsible for the $n_{e}^\mathrm{ped}$ saturation and subsequent pedestal degradation present in the unpumped discharges.

\section{Influence of ionization rate on edge profiles}
\label{sec:ionization}

The success of the pump in lowering plasma density requires both a depletion of ionization source \emph{and} responsiveness in the plasma density to a change in source. The ionization rate, $S_\mathrm{ion}=\langle\sigma v\rangle_\mathrm{ion}n_{e}n_{0}$, depends not only on the density of neutrals, $n_{0}$, but also that of electrons, $n_{e}$. It also depends on the ionization cross-section, $\langle\sigma v\rangle_{ion}$, which is itself a function of both $n_{e}$ and electron temperature, $T_{e}$ \cite{rosenthal_inference_2023}. Since the plasma profiles themselves depend on the ionization source, it is not clear that a change in the density of neutrals alone should directly change the ionization rate and by the same amount. Figure \ref{fig:p0_Sion} quantifies the ability of the pump to modify the ionization source at the separatrix. It plots $S_\mathrm{ion}^\mathrm{sep}$ against the neutral pressures in the main chamber, near the cryopump (cryo), at the OMP, and near the lower divertor (div). The figure shows that for the unpumped discharges, $S_\mathrm{ion}^\mathrm{sep}$ follows the neutral pressure measurements very closely, but there is little trend in the pumped discharges. Generally, neutral pressures near the cryopump and divertor are an order of magnitude larger than at the midplane. When pumping, $p_{0}^\mathrm{OMP}$ and $p_{0}^\mathrm{div}$ decrease by a factor of $\sim 2 - 5$, while $p_{0}^\mathrm{cryo}$ drops by about an order of magnitude.

\begin{figure}
\centering
\includegraphics[width=0.5\columnwidth]{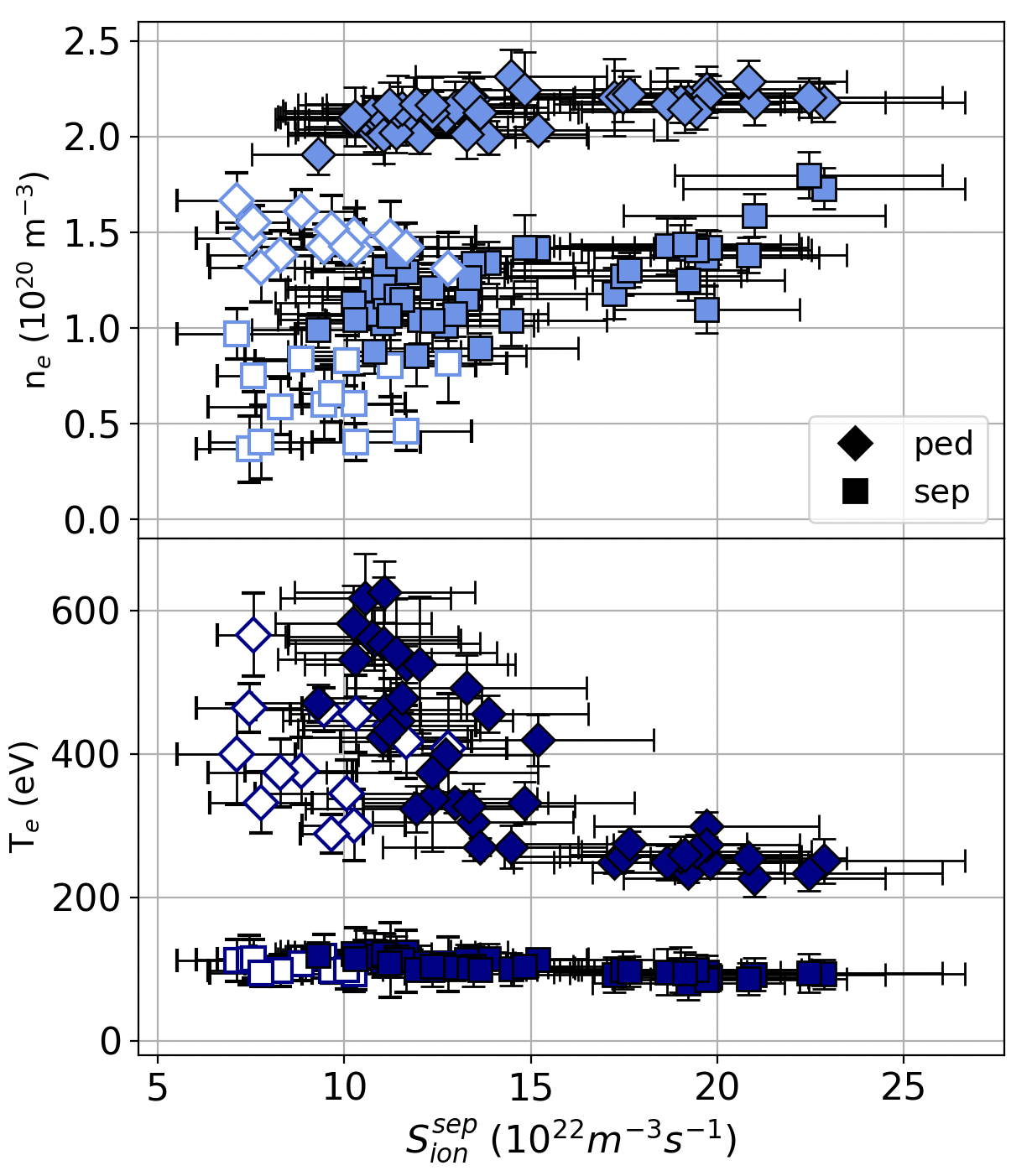}
\caption{Electron density (top) and temperature (bottom) at the pedestal top (diamonds) and separatrix (squares) against the separatrix ionization rate, for both pumped (open) and unpumped (closed) discharges.}
\label{fig:Sion_plasma}
\end{figure}

Figure \ref{fig:Sion_plasma} shows this effect more clearly. The top panel shows $n_{e}^\mathrm{ped}$ and $n_{e}^\mathrm{sep}$ as a function of $S_\mathrm{ion}^\mathrm{sep}$. $n_{e}^\mathrm{sep}$ tracks $S_\mathrm{ion}^\mathrm{sep}$ very closely for both unpumped and pumped discharges. As ionization gets depleted via the pump, so does $n_{e}^\mathrm{sep}$. As with $S_\mathrm{ion}^\mathrm{sep}$, $n_{e}^\mathrm{sep}$ changes only by a factor of five, in contrast with $p_{0}$ which can change by more than an order of magnitude everywhere in the chamber, and up to two orders of magnitude for $p_{0}^\mathrm{cryo}$. Figure \ref{fig:Sion_plasma} also shows $T_{e}^\mathrm{sep}$ as the ionization rate changes. As the ionization rate gets very high, there is a slight depression in $T_{e}^\mathrm{sep}$, but across the dataset, $T_{e}^\mathrm{sep}$ only varies by $\sim$35\%.

At the pedestal top, the dependence of $n_{e}$ and $T_{e}$ on $S_\mathrm{ion}$ is quite different. Figure \ref{fig:Sion_plasma} shows that $n_{e}^\mathrm{ped}$ is not nearly as well-correlated with $S_\mathrm{ion}^\mathrm{sep}$. Though not shown here, it is similarly uncorrelated when plotted against $S_\mathrm{ion}$ at the same location, $S_\mathrm{ion}^\mathrm{ped}$. At high $S_\mathrm{ion}$, $n_{e}^\mathrm{ped}$ is particularly insensitive to increased ionization. At low $S_\mathrm{ion}$, there is a drop in $n_{e}^\mathrm{ped}$, but it does not respond continuously with $S_\mathrm{ion}$. The pedestal density drops by about about 25\%, despite a rather modest change in both $S_\mathrm{ion}^\mathrm{sep}$ and $S_\mathrm{ion}^\mathrm{ped}$. In particular, at $S_\mathrm{ion}^\mathrm{sep} \approx 10^{21}$ m$^{-3}$s$^{-1}$, $n_{e}^\mathrm{ped} \approx$ 2.1 $\times 10^{20}$ m$^{-3}$ when not pumping, and $\approx$ 1.5 $\times 10^{20}$ m$^{-3}$ when pumping. $T_{e}^\mathrm{ped}$ on the other hand, is much more clearly affected by $S_\mathrm{ion}$, at least for $S_\mathrm{ion}^\mathrm{sep} >$ 15 $\times 10^{22}$ m$^{-3}$s$^{-1}$. There, large
increase in ionization is accompanied by a large drop in $T_{e}$. As the pump succeeds in lowering $S_\mathrm{ion}$, higher $T_{e}$ can be sustained at a similarly low level of $P_\mathrm{net}$. This is correlated also with $\nabla T_{e}$ and $\nabla p_{e}$, which helps to explain the higher $H_{98}$ found at lower $P_\mathrm{net}$ seen in Figure \ref{fig:jerry_pnet} when pumping.

\section{The operational space at the top of the pedestal}
\label{sec:os}

\begin{figure}[h!]
\centering
\includegraphics[width=0.5\columnwidth]{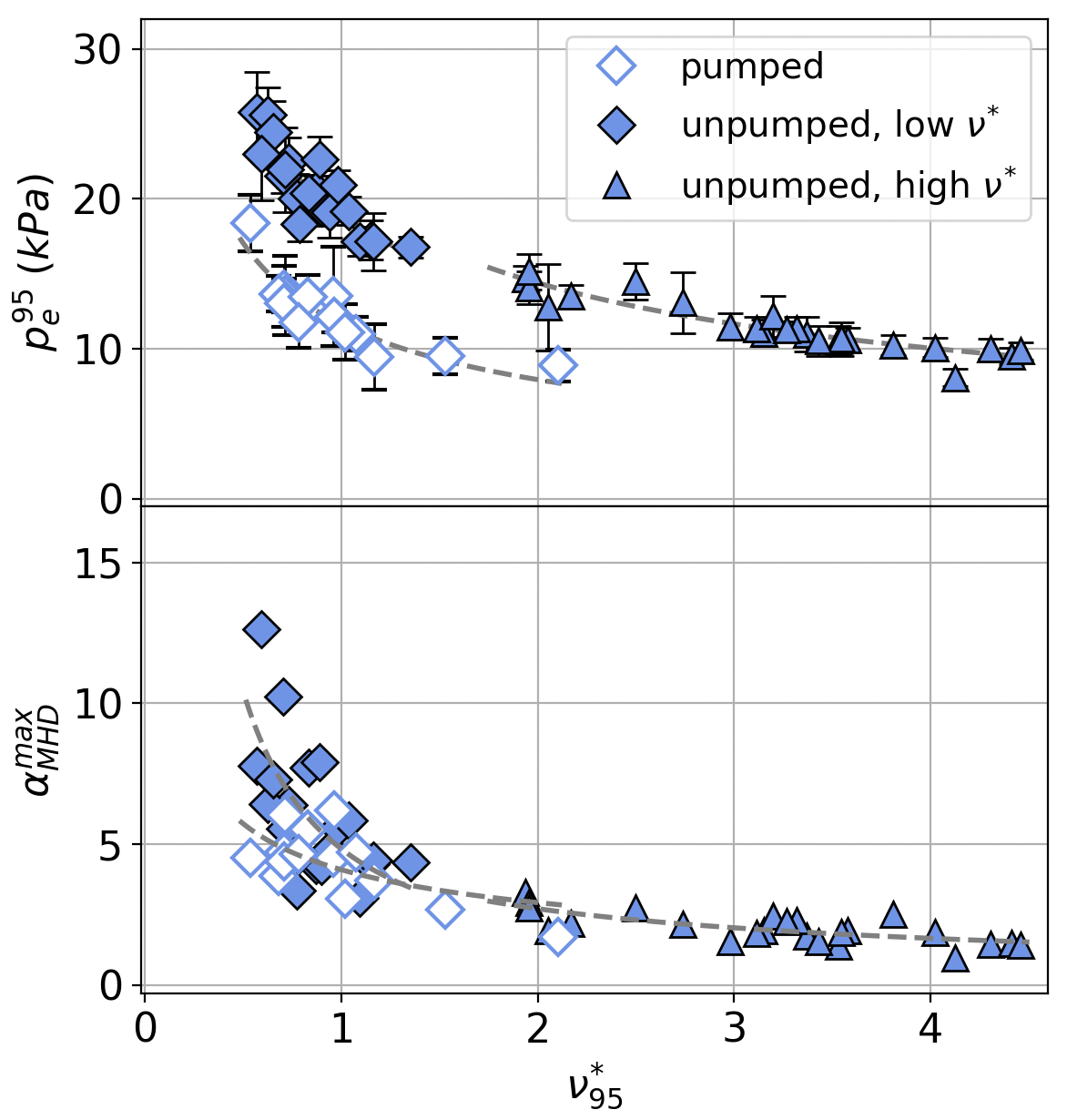}
\caption{Pressure at $\psi_{n} = 0.95$ (top) and $\alpha_\mathrm{MHD}$ (bottom) computed with the maximum $\nabla p_{e}$ against collisionality at $\psi_{n} = 0.95$. A distinction is made for unpumped discharges at $\nu^{*}_{95} > 2$ (closed triangles) and those at $\nu^{*}_{95} < 2$. Dashed black curves are power law regressions $y = C \times (\nu^{*}_{95})^{k}$, shown for each of these ``groups" of discharges.}
\label{fig:nu95_pressures}
\end{figure}

Figure \ref{fig:Sion_plasma} shows that despite rather continuous changes in both ionization and plasma profiles at the separatrix, the pedestal changes are more discrete. To probe this apparent difference, quantities typically associated with pedestal quality are analyzed. The top panel of Figure \ref{fig:nu95_pressures} shows the electron pressure inside the pedestal top, at $\psi_{n}$ = 0.95, $p_{e}^{95}$, plotted against the collisionality at the same location, $\nu^{*}_{95}$, for both sets of plasmas. The latter is calculated using $\nu^{*} = \frac{q_{95}R_{0}\nu_{ei}}{\epsilon^{3/2}v_{\mathrm{th},e}}$. Here $q_{95}$ is the safety factor at $\psi_{n} = 0.95$, $R_{0}$ is the plasma major radius, $\nu_{ei}$ is the electron-ion collision frequency, $\epsilon$ is the inverse aspect ratio, and $v_{\mathrm{th},e}$ is the electron thermal velocity. $\nu_{ei}$ depends on $Z_\mathrm{eff}$, the effective charge of the plasma. For simplicity, $Z_\mathrm{eff}$ is chosen to be 1.4, a common assumption for C-Mod plasmas with no impurity seeding. Since these plasmas are all at similar $B_{t}$ and $I_{P}$, there is little variation in $q_{95}$.

As with $n_{e}^\mathrm{ped}$, the figure shows fairly clear separation between the discharges at the top of the pedestal. The unpumped points can generally be categorized into a low $\nu^{*}$, high $p_{e}$ category (solid diamonds), and a low $\nu^{*}$, high $p_{e}$ category (solid triangles). The pumped discharges then represent a third pedestal condition, lower in both $p_{e}$ \emph{and} $\nu^{*}$. In fact, they have $\nu^{*}_{95}$ similar to that of the low $\nu^{*}$ unpumped points, but $p_{e}^{95}$ similar to that of the high $\nu^{*}$ unpumped points. When each of these three groups of points is regressed with a power law, $p_{e}^{95} = C \times (\nu^{*}_{95})^{k}$, the same exponent, $k = -0.5$ emerges for all three groups of points. Though they have different $a$, $\nu^{*}_{95}$ acts to reduce $p_{e}^{95}$ in the exact same way for all of these plasmas. Alternatively, $p_{e}^{95} = C \times (\nu^{*}_{95})^{-0.5}$ implies that $n_{e}^{95} = C^{2/3}$. Changes to $\nu^{*}_{95}$ are then purely changes to $T_{e}^{95}$, and gradients adjust in such a way as to keep $n_{e}^{95}$ constant despite changes to $\nu^{*}_{95}$ and ionization.

To understand the physical mechanisms that determine the values of $\psi_{n} = 0.95$, inside the pedestal top, it is important to consider pedestal gradients. Previous pedestal research on C-Mod has considered how dimensionless parameters play key roles in regulating pedestal profiles \cite{hughes_edge_2007}. The normalized pressure gradient, $\alpha_\mathrm{MHD}$, is used here as well and tracked also as a function of $\nu^{*}_{95}$. Here, $\alpha_\mathrm{MHD} = \frac{2\mu_{0}q_{95}^{2}R_{0}}{B_{t}^{2}}\nabla p$. Figure \ref{fig:nu95_pressures} shows that as with the pressure, the normalized pressure gradient is very tightly regulated by $\nu^{*}_{95}$. Despite different pressures at the top of the pedestal, the curves relating $\alpha_\mathrm{MHD}$ and $\nu^{*}_{95}$ converge. Though the regression exponent, $b$, varies between the groups of points, the coefficient, $a$ remains between 4 - 5 for all groups.  At low $\nu^{*}_{95}$, there is a rapid drop in $\alpha_\mathrm{MHD}$, which flattens out at high $\nu^{*}_{95}$. All of these pedestals, at relatively fixed $q_{95}$, appear to be governed by dimensionless quantities. They are pinned to a particular normalized pressure gradient, $\alpha_\mathrm{MHD}$, the value of which is given by the pedestal collisionality, $\nu^{*}_{95}$. Variation in both heating and fueling sources (which have been shown to themselves be tightly linked \cite{miller_collisionality_2024}) cause movements along the $\alpha_\mathrm{MHD} - \nu^{*}_{95}$ curve, but do not shift it up and down. This holds even as the power crossing the separatrix and the ionization at the separatrix, $P_\mathrm{net}$ and $S_\mathrm{ion}^\mathrm{sep}$, vary by a factors of 3 - 4.

\section{Interpretive modeling with SOLPS-ITER}
\label{sec:solps}
Figure \ref{fig:p0_Sion} implies that the effect of removing neutrals through pumping on $S_\mathrm{ion}$ and $n_{e}$ at the separatrix at the OMP is at least a 2D problem. SOLPS-ITER simulations are thus performed to supplement the 0D and 1D measurements discussed in Sections \ref{sec:experiment} and \ref{sec:ionization} and further study how sources of particles and heat interact to build up plasma gradients from the separatrix up to the pedestal. This is done for four discharges, at two distinct powers, each with one discharge pumped and one unpumped. Details of the selected discharges are given in Table \ref{tab:solps_selections}.

SOLPS-ITER is used most typically to model the collisional boundary of plasmas, where plasma and neutral-interactions are very important. It includes both a a 2D multi-fluid plasma transport code, called B2.5, and a 3D kinetic Monte Carlo neutral transport solver, EIRENE \cite{wiesen_new_2015}. Given the availability of measurements of both $n_{e}$ and the atomic neutral deuterium, D, density, $n_{0}$, throughout the edge, it is possible to constrain both B2.5 and EIRENE simultaneously. Details of how  both $n_{e}$ and $n_{0}$ are constrained using ETS and LYMID are described in detail in \cite{miller_collisionality_2024}. Of particular importance to this work, however, is that SOLPS-ITER requires user-specified cross-field transport coefficients. For these simulations, a diffusive particle transport coefficient, and electron and ion thermal transport coefficients are specified ($D$, $\chi_{e}$, and $\chi_{i}$ respectively). Important also are the far-SOL boundary conditions. These simulations use leakage boundary conditions (BC) for all channels - density, electron temperature and ion temperature ($\alpha_{n}$, $\alpha_{T_{e}}$, and $\alpha_{T_{i}}$, respectively). This type of BC has been used routinely on C-Mod SOLPS-ITER simulations \cite{reksoatmodjo_role_2021, dekeyser_solps-iter_2017}. Furthermore, it was found to more accurately reproduce the magnitude and allow tuning to the structure of $n_{0}(r)$ across the pedestal.

For simulations of C-Mod plasmas without pumping, no external injection or removal of particles via gas puffing or pumping is needed. Steady-state is achieved with zero particle flux core BC and with an albedo, $R$, set to unity, i.e. perfect reflection, at all material surfaces in the simulation domain. To simulate pumping, however, particle balance is more challenging. Simulating a pump in SOLPS-ITER typically involves reducing $R$ at the designated pumping surface below unity \cite{kukushkin_effect_2007, kaveeva_solps-iter_2023}. In the current simulations, ``pumping" is modeled by designating a pumping surface at the top of a volume constructed to resemble the pump throat, placed below the cryopump volume shown in turquoise in Figure \ref{fig:xs_pump}. $R$ at this surface is reduced to 0.99 (99\% recycling) and particles are removed until the $n_{0}$ throughout the pedestal reaches the level experimentally inferred. This quantity is chosen as the experimentally-relevant figure of merit as opposed to $p_{0}^\mathrm{cryo}$, for example, due to difficulties in calculating a neutral pressure consistent with measurements taken far from the simulated vacuum region (as was the case for C-Mod pressure gauges). In C-Mod experiments with pumping, a steady-state could be reached if the walls begin to outgas (i.e. $R > 1$), or if after some time, the flux of particles to the pump duct becomes negligible. Rather than injecting particles at some other unknown location to balance the particles removed via pumping, once $n_{0}$ reaches the experimentally-observed level, $R$ is returned to unity at all surfaces to halt removal of particles.

In addition to matching the magnitude of $n_{0}$ in the pedestal, focus is also placed on matching the details of neutral penetration at the OMP, $L_{n_{0}}^\mathrm{OMP}$, given by the e-folding length of $n_{0}(r)$ in the region inside the separatrix. Changing $D(r)$ to match $n_{e}(r)$ modifies the particle flux inside the separatrix, which redistributes neutrals and modifies $n_{0}(r)$. At this point, $\alpha_{n}$ and $\alpha_{T_{e,i}}$ are employed to improve the match to both $L_{n_{0}}^\mathrm{OMP}$. The final values and particle content required for the complete match to experimental profiles are tabulated in Table \ref{tab:solps_selections}.

\begin{table}
\begin{center}
\caption{Selected experimental discharges for SOLPS-ITER modeling at two distinct values of $P_{net}$ for each pumping state and final parameters chosen for and resulting from iterative procedure needed to achieve match to experimental profiles.}
\label{tab:solps_selections}
\begin{tabular}{c|c|c|c|c}
\hline
\makecell{Shot \\ number}  & \makecell{$P_\mathrm{net}$ \\(MW)} & Pump & \makecell{$\alpha_{n} / \alpha_{T_{e,i}}$ \\ $(10^{-3})$} & \makecell{Particle \\ content \\ $(10^{20})$} \\
\hline
\hline
1070821004 & 2.0 & off & 2.5 / 2 & 1.2 \\
1080130009 & 2.0 & on & 2.5 / 10 & 0.73 \\
1070821009 & 2.9 & off & 2 / 2 & 1.2 \\
1080125011 & 2.9 & on & 2 / 10 & 0.76 \\
\hline
\end{tabular}
\end{center}
\end{table}

\section{Pedestal transport and the poloidal distribution of neutrals}
\label{sec:poloidal_transport}

\begin{figure}[h!]
\centering
\includegraphics[width=0.7\columnwidth]{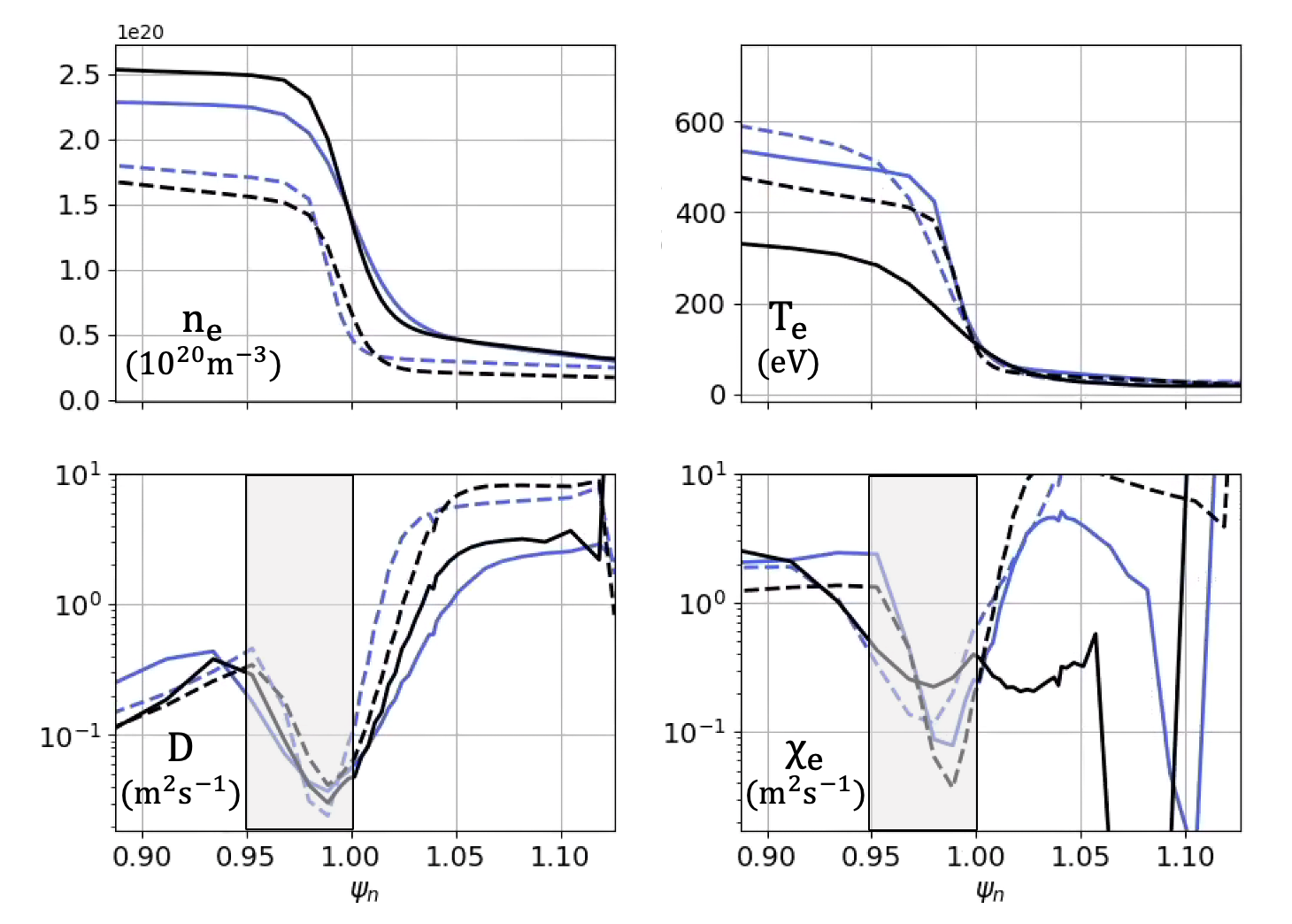}
\caption{Plasma kinetic profiles, $n_{e}$ and $T_{e}$ (top left and top right, respectively), and the transport coefficients for each, $D$ and $\chi_{e}$ (bottom left and bottom right, respectively), resulting from SOLPS-ITER simulations. Blue/black curves are at high/low $P_\mathrm{net}$ and solid/dashed curves are unpumped/pumped.}
\label{fig:solps_plasma}
\end{figure}

Figure \ref{fig:solps_plasma} shows the resulting experimentally-matched plasma profiles for the four cases, as well as the transport coefficient profiles required to achieve the match outlined above. Given that special attention was placed on matching $n_{0}$ only inside the separatrix, analysis of the transport coefficients is restricted to the steep-gradient pedestal region. In particular, the minimum value of these coefficients, which is generally co-located with the maximum gradient in the respective kinetic profiles, is tracked. While changes in the magnitude of $D$ appear small, differences, in particular with experimentally-determined $D_\mathrm{eff}$, are important. These are explained in greater detail below. There is also a small decrease in the width of the $D$ transport well for the pumped discharges. Analysis of $\chi_{e}$ and its shape are not within the scope of the current work.

\begin{table}
\footnotesize
\begin{center}
\caption{Transport coefficients, inferred experimentally and from SOLPS simulations, as well as the ratio between the simulated and experimental inference, and the ratio of OMP to LXP neutral density as a figure of merit for the poloidal asymmetry of neutrals}
\label{tab:poloidal_transport}
\begin{tabular}{c|c|c|c|c|c|c}
\hline
Shot \# & Simulation type & \makecell{$D_\mathrm{eff}^\mathrm{exp}$ \\ (10$^{-2}$ m$^{2}$s$^{-1}$)} & \makecell{$D_\mathrm{eff}^\mathrm{SOLPS}$ \\ (10$^{-2}$ m$^{2}$s$^{-1}$)} & \makecell{$\chi_{e}^\mathrm{SOLPS}$ \\ (10$^{-2}$ m$^{2}$s$^{-1}$)} & $\frac{D_\mathrm{eff}^\mathrm{SOLPS}}{D_\mathrm{eff}^\mathrm{exp}}$ & $\frac{n_{0}^\mathrm{OMP}}{n_{0}^\mathrm{LXP}}$  \\
\hline
\hline
1070821004 & low $P_\mathrm{net}$, unpumped & 4.3 & 3.0 & 20 & 0.70 & 10.2 \\
1080130009 & low $P_\mathrm{net}$, pumped & 1.4 & 4.1 & 3.7 & 2.9 & 0.67 \\
1070821009 & high $P_\mathrm{net}$, unpumped & 4.9 & 3.7 & 7.8 & 0.76 & 10.9 \\
1080125011 & high $P_\mathrm{net}$, pumped & 0.56 & 2.4 & 12 & 4.3 & 0.69 \\
\hline
\end{tabular}
\end{center}
\end{table}

The 3nd and 4th columns of Table \ref{tab:poloidal_transport} show the experimentally determined values for $D_\mathrm{eff}^\mathrm{exp}$ at mid-pedestal, as well as those determined from the SOLPS simulation, $D_\mathrm{eff}^\mathrm{SOLPS}$. Profiles of the experimental diffusivity are calcualated using $D_\mathrm{eff} = \frac{\Gamma_{D}}{-\nabla n_{e}}$, where $\Gamma_{D}$ is the cross-field particle flux, computed from the integral of $S_\mathrm{ion}$ and $\nabla n_{e}$ is computed analytically from the assumed modified hyperbolic tangent functional fit form for both $n_{e}$ and $T_{e}$. Details of how this coefficient is calculated for LYMID and ETS data can be found in  \cite{miller_collisionality_2024}. Comparing experimental measurements alone shows that for constant $P_\mathrm{net}$, a change in $D_\mathrm{eff}$ occurs as plasmas are pumped down.

\begin{figure}
\centering
\includegraphics[width=1\columnwidth]{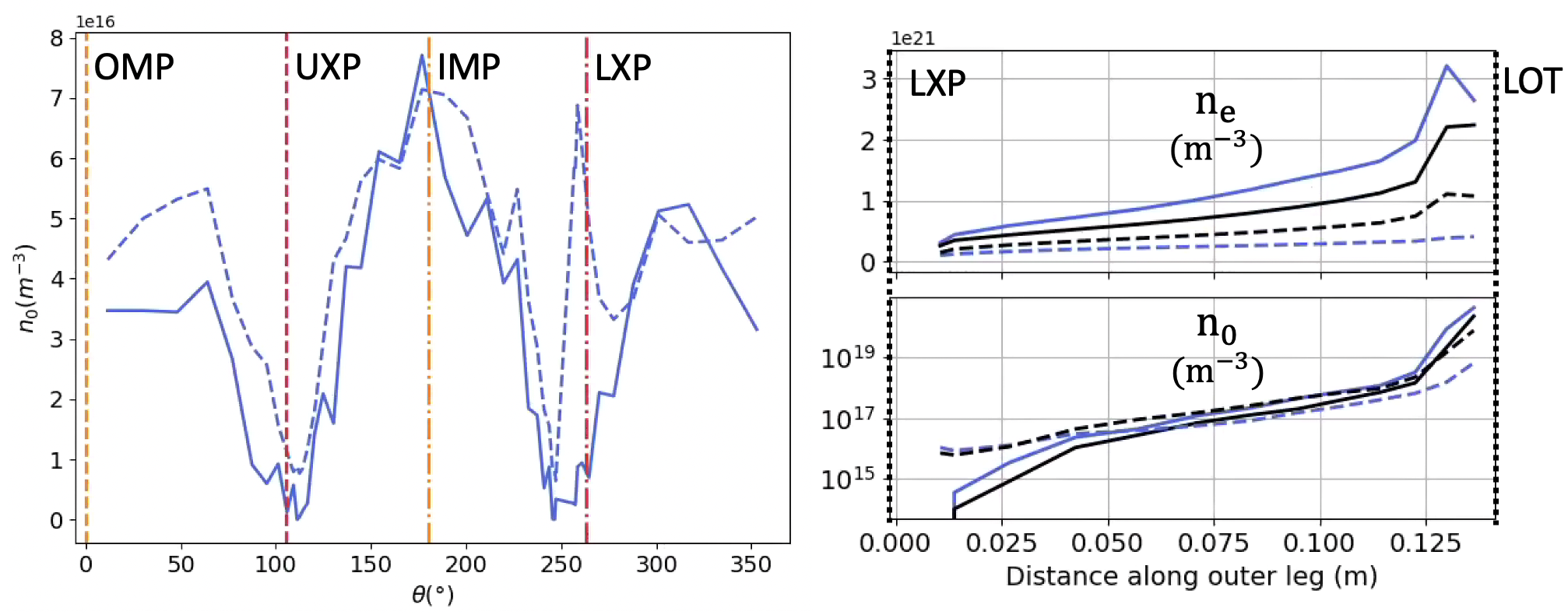}
\caption{Atomic neutral density plotted against poloidal angle (left) and profiles along outer leg for $n_{e}$ (top right) and $n_{0}$ (bottom right). Left panel only shows the pumped (dashed) and unpumped (solid) cases at lower $P_\mathrm{net}$. Poloidal angle uses $\theta = 0\deg$ at the OMP and increases counter-clockwise along the poloidal cut. Vertical lines denote poloidal positions of interest (shown in first $n_{0}$ contour plot at right). Midplanes are shown in yellow, X-points in red. Dashed lines are OMP and upper X-point (UXP). Dash-dotted lines are the inner midplane (IMP) and lower X-point (LXP). Right panel shows profiles for same cases as well as for pumped and unpumped cases at higher $P_\mathrm{net}$. For right panels, distance along outer leg increases away from the LXP and towards the lower outer target (LOT). }
\label{fig:poloidal_pen}
\end{figure}

Results from SOLPS simulations confirm that for both unpumped discharges, $D_\mathrm{eff}^\mathrm{exp} \approx D_\mathrm{eff}^\mathrm{SOLPS}$, i.e. the 1D plasma and neutral profiles are sufficient for describing particle transport in the pedestal. For both pumped discharges, $D_\mathrm{eff}^\mathrm{exp} < D_\mathrm{eff}^\mathrm{SOLPS}$, i.e. the 1D measurements under-predict the simulated $D_\mathrm{eff}$. To understand why the experimentally inferred $D_\mathrm{eff}$ under-predicts the value that results from the SOLPS simulations, it is important to probe the 2D distribution of neutrals and how pumping changes this. Variation in the poloidal distribution of neutrals across the pedestal will affect the distribution of plasma particle fluxes. Differences in particle flux at the OMP imply differences in the $D_\mathrm{eff}$ required to sustain the $\nabla n_{e}$ that is experimentally observed. The left panel of Figure \ref{fig:poloidal_pen} shows the poloidal distributions of neutrals at fixed radial location, the separatrix, as a function of poloidal angle, $\theta$. To reduce visual clutter, only the pumped and unpumped discharges at high $P_\mathrm{net}$ are shown. Results for low $P_\mathrm{net}$ are very similar, as indicated in the last column of Table \ref{tab:poloidal_transport}. The middle and right panels of the same figure show the contours of $n_{0}$ for both high $P_\mathrm{net}$ cases.

The left panel of Figure \ref{fig:poloidal_pen} shows that $n_{0}$ peaks at the midplanes in both discharges. This is in line with experimental measurements in ohmic discharges on Alcator C-Mod, as well as simulations of these ohmic plasmas and H-mode plasmas using UEDGE and SOLPS-ITER respectively \cite{labombard_cross-field_2000, reksoatmodjo_role_2021}. These results indicate that in Alcator C-Mod, the majority of neutrals recycle off main-chamber limiters and enter the confined region at the midplanes, rather than at the X-points. The reason for this is not fully known, but it is thought to be related to high cross-field particle fluxes to main chamber walls, concurrent with density shoulder formation \cite{labombard_cross-field_2000}, a situation in which $n_{e}$ remains high well into the SOL. While the pumped discharge does still experience accumulation of neutrals near both the inner and the outer midplane, $n_{0}$ also becomes large near the lower X-point (LXP). This is in stark contrast to the unpumped discharge, where a negligible number of neutrals appears at this poloidal location.

To study poloidal neutral penetration, it is important to consider the divertor plasma. The right panels of Figure \ref{fig:poloidal_pen} show $n_{e}$ and $n_{0}$ along the lower outer divertor leg. The bottom panel of Figure \ref{fig:poloidal_pen} shows $n_{0}$ as a function of the distance from the X-point along the outer divertor leg, for all four discharges. In all four cases, $n_{0}$ peaks at the lower outer target (LOT), where recycling is high, and decays as it approaches the LXP. The decay length along the leg, $L_{n_{0}}$, varies substantially between the pumped discharges and the unpumped discharges. For the unpumped discharges, $n_{0}$ is larger at the outer target than for the pumped discharges, but decays more quickly away from the target. In contrast, the pumped discharges have lower $n_{0}$ at the target, but longer penetration lengths. By the time neutrals arrive at the LXP, $n_{0}$ reaches a value comparable to that at the OMP. The top panel of Figure \ref{fig:poloidal_pen} also shows $n_{e}$ along the divertor leg. For both pumped discharges, $n_{e}$ is lower than in the unpumped discharges along the whole leg. $n_{e}$ has been previously linked to neutral opaqueness, whereby $L_{n_{0}}$ is related inversely with $n_{e}$ in the pedestal \cite{hughes_advances_2006, mordijck_overview_2020}. A similar relationship appears here, but for the SOL plasma. Lower $n_{e}$ along the divertor leg leads to larger $L_{n_{0}}$ for pumped discharges.

Poloidally-varying $L_{n_{0}}$ has been previously observed, also in SOLPS-ITER simulations of C-Mod \cite{reksoatmodjo_role_2021}. As suggested above, an off-midplane accumulation of neutrals implies also a change to the flux-surface averaged particle flux, which is what ultimately dictates $D_\mathrm{eff}^\mathrm{SOLPS}$. As $n_{0}$ near the X-point approaches that at the midplanes, the plasma particle flux that is needed to balance the additional influx of neutrals must also increase. Thus, to achieve the same $\nabla n_{e}^\mathrm{ped}$, a larger $D_\mathrm{eff}$ is required. Table \ref{tab:poloidal_transport} shows the ratio of $n_{0}$ at the OMP, $n_{0}^\mathrm{OMP}$, to that at the lower X-point, $n_{0}^\mathrm{LXP}$, for the four different plasmas. For all unpumped discharges, $\frac{n_{0}^\mathrm{OMP}}{n_{0}^\mathrm{LXP}} >> 1$, while for all pumped discharges, $\frac{n_{0}^\mathrm{OMP}}{n_{0}^\mathrm{LXP}} \sim 1$, The same table also shows the ratio, $\frac{D_\mathrm{eff}^\mathrm{SOLPS}}{D_\mathrm{eff}^\mathrm{exp}}$. As noted earlier, this ratio is close to one in the unpumped discharges, when $\frac{n_{0}^\mathrm{OMP}}{n_{0}^\mathrm{LXP}} >> 1$, but is greater than one when $\frac{n_{0}^\mathrm{OMP}}{n_{0}^\mathrm{LXP}} \sim 1$. This number, essentially a comment on the most important fueling location, proves a good proxy for the relative ratio between flux-surface averaged particle transport coefficient evaluated with a 2D source, compared to one evaluated simply from 1D profiles, in this case at the OMP.

\section{Conclusions and discussion}
\label{sec:conclusions}

This work supplements experimental measurements with simulations to examine the influence of particle control via a cryopump on the edge plasma profiles of Alcator C-Mod. It begins by probing changes to $p_{0}$ induced by pumping and how this propagates to changes to the $S_\mathrm{ion}$ at the OMP. $S_\mathrm{ion}^\mathrm{OMP}$ is generally lower when pumping, but its relationship with $p_{0}$ at the cryopump is non-monotonic. While pumping is found to substantially influence neutral pressures throughout the vessel, it is found to only slightly lower $S_\mathrm{ion}$. Experimentally, $n_{e}^\mathrm{sep}$ is found to be sensitive to changes to $S_\mathrm{ion}$, as is $T_{e}^\mathrm{ped}$, although also not as sensitive as $p_{0}$. While $n_{e}^\mathrm{ped}$ does vary somewhat in this dataset, it appears easily controllable through source modification. Instead, consideration of $p_{e}$, $\nu^{*}$, and $\alpha_\mathrm{MHD}$ shows that changes to the sources modify $\nu^{*}_{95}$, and thus $\alpha_\mathrm{MHD}$, in such a way to keep $n_{e}^\mathrm{ped}$ relatively fixed. The data provide evidence that plasma transport is dominant in setting profiles. 

To better understand how particle and heat sources couple to transport and what role 2D effects might have on upstream sources, SOLPS-ITER simulations are performed. The simulations confirm that modifications to both $P_\mathrm{net}$ and $S_\mathrm{ion}^\mathrm{sep}$ \emph{also} require different $D$ and $\chi_{e}$ coefficients to achieve matches to plasma and neutral profiles, providing some evidence of non-diffusive transport. Changes to $D$ and $\chi_{e}$ with source fluxes, whether particle or heat, is a feature of profiles held at critical gradients, not a diffusive system. 2D effects resulting from poloidally-asymmetric fueling are found to be important in contributing to $\Gamma_{D}$ and $D$. These asymmetries suggest that $\Gamma_{D}$, and thus, $S_\mathrm{ion}^\mathrm{sep}$, might remain high even when the neutral source is depleted at one location if it means the redistribution of neutrals to another. A fully self-consistent model for $S_\mathrm{ion}^\mathrm{sep}$, and eventually for the flux-gradient relationship in the plasma edge, might also require some information about poloidal asymmetries.

Modification of the neutral distribution via cryopumping and a decrease in the opaqueness of the divertor legs is one way in which the ratio $\frac{n_{0}^\mathrm{OMP}}{n_{0}^\mathrm{LXP}}$ might change. This ratio might also change in the same direction if the numerator, $n_{0}^\mathrm{OMP}$, decreases. As noted earlier, the accumulation of neutrals at the midplanes has been tied to large $\Gamma_{D}$ at main chamber limiter surfaces. Were $\Gamma_{D}^\mathrm{SOL}$ lowered via decreased perpendicular SOL transport, for example, there might also be a considerable drop in $n_{0}^\mathrm{OMP}$. The ratio $\frac{n_{0}^\mathrm{OMP}}{n_{0}^\mathrm{LXP}}$ might then change independently of neutral opaqueness in the divertor, the effect identified here. The formation of a density shoulder, however, is tied to large values of Greenwald fraction, $f_{G}$ \cite{labombard_evidence_2005}. Higher $f_{G}$ at fixed $I_{P}$ implies larger $n_{e}^\mathrm{div}$, which would further increase neutral opaqueness along the divertor legs, as well as $\frac{n_{0}^\mathrm{OMP}}{n_{0}^\mathrm{LXP}}$. A greater understanding of the interplay between cross-field particle transport and fueling across the separatrix and at the divertor is necessary to understand what mechanisms determine sources and transport in the pedestal and at the separatrix.

\section*{Acknowledgements}
\noindent The authors would like to thank A.M. Rosenthal, F. Sciortino, and T. Odstrcil for their assistance with data analysis. They would also like to thank R. Masline and Y.C. Chuang for their support with simulation. This material is based upon work supported by the U.S. Department of Energy, Office of Science, Office of Fusion Energy Sciences, under Awards DE-SC0021629 and DE-SC0007880, and the MIT Presidential Fellowship.

\bibliographystyle{unsrt}
\bibliography{references}% Produces the bibliography via BibTeX.

\begin{thebibliography}{10}

\bibitem{houlberg_density_1994}
W.A Houlberg, S.E Attenberger, and M.J Grapperhaus.
\newblock Density profile control in a fusion reactor using pellet injection.
\newblock {\em Nuclear Fusion}, 34(1):93--108, January 1994.

\bibitem{lawson_criteria_1957}
J~D Lawson.
\newblock Some {Criteria} for a {Power} {Producing} {Thermonuclear} {Reactor}.
\newblock {\em Proceedings of the Physical Society. Section B}, 70(1):6--10, January 1957.

\bibitem{asdex_team_h-mode_1989}
{ASDEX Team}.
\newblock The {H}-{Mode} of {ASDEX}.
\newblock {\em Nuclear Fusion}, 29(11):1959--2040, November 1989.

\bibitem{burrell_effects_1997}
K.~H. Burrell.
\newblock Effects of {E}×{B} velocity shear and magnetic shear on turbulence and transport in magnetic confinement devices.
\newblock {\em Physics of Plasmas}, 4(5):1499--1518, May 1997.

\bibitem{kotschenreuther_quantitative_1995}
M.~Kotschenreuther, W.~Dorland, M.~A. Beer, and G.~W. Hammett.
\newblock Quantitative predictions of tokamak energy confinement from first-principles simulations with kinetic effects.
\newblock {\em Physics of Plasmas}, 2(6):2381--2389, June 1995.

\bibitem{greenwald_h_1997}
M~Greenwald, R.L Boivin, F~Bombarda, P.T Bonoli, C.L Fiore, D~Garnier, J.A Goetz, S.N Golovato, M.A Graf, R.S Granetz, S~Horne, A~Hubbard, I.H Hutchinson, J.H Irby, B~LaBombard, B~Lipschultz, E.S Marmar, M.J May, G.M McCracken, P~O'Shea, J.E Rice, J~Schachter, J.A Snipes, P.C Stek, Y~Takase, J.L Terry, Y~Wang, R~Watterson, B~Welch, and S.M Wolfe.
\newblock H mode confinement in {Alcator} {C}-{Mod}.
\newblock {\em Nuclear Fusion}, 37(6):793--807, June 1997.

\bibitem{kinsey_iter_2011}
J.E. Kinsey, G.M. Staebler, J.~Candy, R.E. Waltz, and R.V. Budny.
\newblock {ITER} predictions using the {GYRO} verified and experimentally validated trapped gyro-{Landau} fluid transport model.
\newblock {\em Nuclear Fusion}, 51(8):083001, August 2011.

\bibitem{frassinetti_global_2017}
L.~Frassinetti, M.N.A. Beurskens, S.~Saarelma, J.E. Boom, E.~Delabie, J.~Flanagan, M.~Kempenaars, C.~Giroud, P.~Lomas, L.~Meneses, C.S. Maggi, S.~Menmuir, I.~Nunes, F.~Rimini, E.~Stefanikova, H.~Urano, and G.~Verdoolaege.
\newblock Global and pedestal confinement and pedestal structure in dimensionless collisionality scans of low-triangularity {H}-mode plasmas in {JET}-{ILW}.
\newblock {\em Nuclear Fusion}, 57(1):016012, January 2017.

\bibitem{rodriguez-fernandez_predictions_2020}
P.~Rodriguez-Fernandez, N.~T. Howard, M.~J. Greenwald, A.~J. Creely, J.~W. Hughes, J.~C. Wright, C.~Holland, Y.~Lin, F.~Sciortino, and {the SPARC team}.
\newblock Predictions of core plasma performance for the {SPARC} tokamak.
\newblock {\em Journal of Plasma Physics}, 86(5):865860503, October 2020.

\bibitem{hughes_pedestal_2013}
J.W. Hughes, P.B. Snyder, J.R. Walk, E.M. Davis, A.~Diallo, B.~LaBombard, S.G. Baek, R.M. Churchill, M.~Greenwald, R.J. Groebner, A.E. Hubbard, B.~Lipschultz, E.S. Marmar, T.~Osborne, M.L. Reinke, J.E. Rice, C.~Theiler, J.~Terry, A.E. White, D.G. Whyte, S.~Wolfe, and X.Q. Xu.
\newblock Pedestal structure and stability in {H}-mode and {I}-mode: a comparative study on {Alcator} {C}-{Mod}.
\newblock {\em Nuclear Fusion}, 53(4):043016, April 2013.

\bibitem{faitsch_analysis_2023}
M.~Faitsch, T.~Eich, G.F. Harrer, E.~Wolfrum, D.~Brida, P.~David, M.~Dunne, L.~Gil, B.~Labit, and U.~Stroth.
\newblock Analysis and expansion of the quasi-continuous exhaust ({QCE}) regime in {ASDEX} {Upgrade}.
\newblock {\em Nuclear Fusion}, 63(7):076013, July 2023.

\bibitem{snyder_characterization_2004}
P~B Snyder, H~R Wilson, T~H Osborne, and A~W Leonard.
\newblock Characterization of peeling–ballooning stability limits on the pedestal.
\newblock {\em Plasma Physics and Controlled Fusion}, 46(5A):A131--A141, May 2004.

\bibitem{kallenbach_neutral_2019}
A.~Kallenbach, M.~Bernert, R.~Dux, T.~Eich, S.S. Henderson, T.~Pütterich, F.~Reimold, V.~Rohde, and H.J. Sun.
\newblock Neutral pressure and separatrix density related models for seed impurity divertor radiation in {ASDEX} {Upgrade}.
\newblock {\em Nuclear Materials and Energy}, 18:166--174, January 2019.

\bibitem{henderson_parameter_2021}
S.S. Henderson, M.~Bernert, C.~Giroud, D.~Brida, M.~Cavedon, P.~David, R.~Dux, J.R. Harrison, A.~Huber, A.~Kallenbach, J.~Karhunen, B.~Lomanowski, G.~Matthews, A.~Meigs, R.A. Pitts, F.~Reimold, M.L. Reinke, S.~Silburn, N.~Vianello, S.~Wiesen, and M.~Wischmeier.
\newblock Parameter dependencies of the experimental nitrogen concentration required for detachment on {ASDEX} {Upgrade} and {JET}.
\newblock {\em Nuclear Materials and Energy}, 28:101000, September 2021.

\bibitem{Moulton_Lengyel_2021}
D.~Moulton, P.C. Stangeby, X.~Bonnin, and R.A. Pitts.
\newblock {Comparison between SOLPS-4.3 and the Lengyel Model for ITER baseline neon-seeded plasmas}.
\newblock {\em Nuclear Fusion}, 61(4):046029, 8 2021.

\bibitem{eich_separatrix_2021}
T~Eich and P~Manz.
\newblock The separatrix operational space of {ASDEX} {Upgrade} due to interchange-drift-{Alfve}´n turbulence.
\newblock {\em Nucl. Fusion}, 2021.

\bibitem{manz_power_2023}
P.~Manz, T.~Eich, and O.~Grover.
\newblock The power dependence of the maximum achievable {H}-mode and (disruptive) {L}-mode separatrix density in {ASDEX} {Upgrade}.
\newblock {\em Nuclear Fusion}, 63(7):076026, July 2023.

\bibitem{brian_pump_2007}
B.~LaBombard, et~al.
\newblock Commissioning of the upper diveror cryopump system in {Alcator C-Mod}.

\bibitem{hughes_high-resolution_2001}
J.~W. Hughes, D.~A. Mossessian, A.~E. Hubbard, E.~S. Marmar, D.~Johnson, and D.~Simon.
\newblock High-resolution edge {Thomson} scattering measurements on the {Alcator} {C}-{Mod} tokamak.
\newblock {\em Review of Scientific Instruments}, 72(1):1107--1110, January 2001.

\bibitem{miller_collisionality_2024}
M.A. Miller, et~al.
\newblock Enhanced pedestal transport driven by edge collisionality on {Alcator C-Mod} and its role in regulating {H-mode} pedestal gradients.
\newblock Submitted, 2024.

\bibitem{rosenthal_inference_2023}
A.M. Rosenthal, J.W. Hughes, F.M. Laggner, T.~Odstrčil, A.~Bortolon, T.M. Wilks, and F.~Sciortino.
\newblock Inference of main ion particle transport coefficients with experimentally constrained neutral ionization during edge localized mode recovery on {DIII}-{D}.
\newblock {\em Nuclear Fusion}, 63(4):042002, April 2023.

\bibitem{hughes_edge_2007}
J.W Hughes, B~LaBombard, J~Terry, A~Hubbard, and B~Lipschultz.
\newblock Edge profile stiffness and insensitivity of the density pedestal to neutral fuelling in {Alcator} {C}-{Mod} edge transport barriers.
\newblock {\em Nuclear Fusion}, 47(8):1057--1063, August 2007.

\bibitem{wiesen_new_2015}
S.~Wiesen, D.~Reiter, V.~Kotov, M.~Baelmans, W.~Dekeyser, A.S. Kukushkin, S.W. Lisgo, R.A. Pitts, V.~Rozhansky, G.~Saibene, I.~Veselova, and S.~Voskoboynikov.
\newblock The new {SOLPS}-{ITER} code package.
\newblock {\em Journal of Nuclear Materials}, 463:480--484, August 2015.

\bibitem{reksoatmodjo_role_2021}
R.~Reksoatmodjo, S.~Mordijck, J.W. Hughes, J.D. Lore, and X.~Bonnin.
\newblock The role of edge fueling in determining the pedestal density in high neutral opacity {Alcator} {C}-{Mod} experiments.
\newblock {\em Nuclear Materials and Energy}, 27:100971, June 2021.

\bibitem{dekeyser_solps-iter_2017}
W.~Dekeyser, X.~Bonnin, S.W. Lisgo, R.A. Pitts, Dan Brunner, Brian LaBombard, and Jim~L. Terry.
\newblock {SOLPS}-{ITER} {Study} of neutral leakage and drift effects on the alcator {C}-{Mod} divertor plasma.
\newblock {\em Nuclear Materials and Energy}, 12:899--907, August 2017.

\bibitem{kukushkin_effect_2007}
A.S. Kukushkin, H.D. Pacher, V.~Kotov, D.~Reiter, D.P. Coster, and G.W. Pacher.
\newblock Effect of the dome on divertor performance in {ITER}.
\newblock {\em Journal of Nuclear Materials}, 363-365:308--313, June 2007.

\bibitem{kaveeva_solps-iter_2023}
E.~Kaveeva, S.~Makarov, I.~Senichenkov, V.~Rozhansky, I.~Veselova, X.~Bonnin, and R.A. Pitts.
\newblock {SOLPS}-{ITER} modeling of deuterium throughput impact on the {ITER} {SOL} plasma.
\newblock {\em Nuclear Materials and Energy}, 35:101424, June 2023.

\bibitem{labombard_cross-field_2000}
B~LaBombard, M.V Umansky, R.L Boivin, J.A Goetz, J~Hughes, B~Lipschultz, D~Mossessian, C.S Pitcher, J.L Terry, and Alcator Group.
\newblock Cross-field plasma transport and main-chamber recycling in diverted plasmas on {Alcator} {C}-{Mod}.
\newblock {\em Nuclear Fusion}, 40(12):2041--2060, December 2000.

\bibitem{hughes_advances_2006}
J.~W. Hughes, B.~LaBombard, D.~A. Mossessian, A.~E. Hubbard, J.~Terry, T.~Biewer, and The Alcator C-Mod Team.
\newblock Advances in measurement and modeling of the high-confinement-mode pedestal on the {Alcator} {C}-{Mod} tokamak.
\newblock {\em Physics of Plasmas}, 13(5):056103, May 2006.

\bibitem{mordijck_overview_2020}
S.~Mordijck.
\newblock Overview of density pedestal structure: role of fueling versus transport.
\newblock {\em Nuclear Fusion}, 60(8):082006, August 2020.

\bibitem{labombard_evidence_2005}
B~LaBombard, J.W Hughes, D~Mossessian, M~Greenwald, B~Lipschultz, J.L Terry, and The Alcator C-Mod Team.
\newblock Evidence for electromagnetic fluid drift turbulence controlling the edge plasma state in the {Alcator} {C}-{Mod} tokamak.
\newblock {\em Nuclear Fusion}, 45(12):1658--1675, December 2005.

\end{thebibliography}

\end{document}